\documentclass[aps,pre,reprint,groupedaddress]{revtex4-1}
\usepackage{graphicx}
\usepackage{dcolumn}
\usepackage{bm}
\usepackage{amsmath}
\usepackage{amsfonts}
\usepackage{subeqnar}
\usepackage{color}

\begin{document}

\title{%
Low-dimensional functionality of complex network dynamics:
Neuro-sensory integration in the Caenorhabditis elegans connectome}

\author{James Kunert$^1$\footnote{Electronic address: \texttt{kunert@uw.edu}}, Eli Shlizerman$^2$ and J. Nathan Kutz$^2$}%
%
\affiliation{$^1$ Department of Physics, University of Washington, Seattle, WA. 98195}

\affiliation{$^2$ Department of Applied Mathematics, University of Washington, Seattle, WA. 98195-2420}

\date{\today}

\begin{abstract}
We develop a biophysical model of neuro-sensory integration in 
the model organism {\em Caenorhabditis elegans}.  Building on
experimental findings of the neuron conductances and
their resolved connectome, we posit the first full dynamic model of the
neural voltage excitations that allows for a characterization of network structures which link input stimuli to neural proxies of behavioral responses. Full connectome simulations of neural responses to prescribed inputs show 
that robust, low-dimensional bifurcation structures drive neural voltage activity modes.
Comparison of these modes with experimental studies allows us to link these network structures to behavioral responses. Thus 
the underlying bifurcation structures discovered, i.e. induced Hopf bifurcations, are critical in explaining behavioral responses such as swimming and crawling.
\end{abstract}


\maketitle
\section{Introduction}
Complex physical systems comprised of a network of nonlinear
dynamical components of voltage activity are capable of producing robust functionality and/or 
low-dimensional patterns of coherent activity.
The coherent swing instability in power grid networks~\cite{Mezic11}, for instance, is an example of these phenomena which
have been observed in experiments and computational studies, yet are difficult to characterize with theoretical techniques.
Other examples of interacting dynamical systems that are well-known in physics, and that
produce functional behavior or coherent patterns, include coupled oscillators (e.g. the Kuramoto 
oscillators),  analog circuits, coupled lasers, many-particle systems, etc.
Biophysical systems, whose interactions are often driven by chemical reactions,
voltage activity, and/or ion exchange, produce similar functionality and structured activity.

\begin{figure}[b]
 \includegraphics[width=1.0\columnwidth]{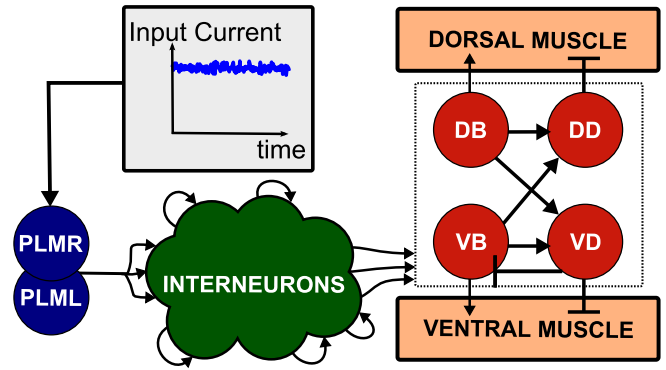}%
 \caption{\label{fig:PLMdiagram}
In our full network simulations, the neurons receive a physically realistic stimulus (input current) which varies in amplitude for different trials.  For the example illustrated, PLM neurons are stimulated leading to signal
propagation through a network of densely connected interneurons which activate the motorneuron subcircuits and low-dimensional neural response modes~\cite{Bialek} 
that control forward locomotion~\cite{Aravi12}. }
 \end{figure}

Neuro-sensory networks, which are an important subclass of biophysical systems, are ideal for characterizing the role of seemingly complex
network interactions for producing robust functionality and can motivate
bio-inspired engineering principles.
Neuro-sensory integration, which attempts to understand the neural pathways from input stimuli to motor-neuron driven behavioral 
responses and low-dimensional movements, is one of the most challenging and open problems in the field of neuroscience today.  
The primary challenge lies in understanding how large networks of different classes of neurons (e.g. sensory-, inter- and motor-neurons which can be either inhibitory or excitatory) 
interact to produce the observed robust behavioral responses to stimuli.  
Ultimately, the biophysical processes produce a large, nonlinear network of electronic conductances that dynamically decode input stimulus and drive downstream neuronal function and behavior
(See, for instance, Fig.~\ref{fig:PLMdiagram}). 

The nematode {\em Caenorhabditis elegans} (\emph{C. elegans}) is a perfect model organism to consider in the context of neuro-sensory integration as it is comprised of only 302 sensory, motor and inter-neurons whose electro-physical connections (i.e. its connectome) are known from serial section electron microscopy~\cite{White,Chen}.  By combining the known connectome data~\cite{Varshney} with a physiologically appropriate neuron model~\cite{Wicks,Shlizerman},
we are able to model the full neural network dynamics in response to time-dependent stimuli.  Such efforts allow for a theoretical characterization of the network biophysics and voltage activity that drives the neuro-sensory integration process in {\em C. elegans} and determine its ability to elicit behavioral responses~\cite{Sharad,Aravi09,Aravi12}.  
Our studies show that input stimuli can produce bifurcations in the neuronal network that
drive low-dimensional responses associated with behavioral activity.  Specifically, we show that
stimulation of the PLM neurons, for instance, induces a Hopf bifurcation that leads to the
onset of a robust two-mode oscillatory behavior associated with crawling.
This is the first study of its kind computationally relating the sensory input with the resultant full-connectome dynamical behavior of the inter- and motor-neurons.

The {\em C. elegans} is an important model organism due to the fact that (i) it possesses only a small number of sensory neurons, often linked to specific stimuli~\cite{WormAtlas}, and (ii) its range of behavioral responses are varied yet limited, confined to swimming, crawling, turning and performing chemotaxis, for instance.   Thus it is reasonable to posit a complete model of its neuro-sensory integration capabilities.  
Aiding in this effort is the near-complete connectivity data for the gap junctions and chemical synapses connecting the sensory neurons to the inter- and motor-neurons~\cite{Varshney}. Moreover, current experiments measure the response of various neurons to input stimuli since a description of these responses cannot be drawn from the static connectivity data alone. These studies suggest that computational modeling can assist in describing neural dynamics and their relation to the connectome. 

\section{Neuron Dynamics}

Simulations of {\em C. elegans} neural dynamics are challenging since (i) it it difficult to measure electrical parameters which characterize precisely the directionality and conductance of each connection, and (ii) the single neuron dynamics do not appear to be characterized by standard spiking neuron models. Indeed, genomic sequencing and electro-physiological studies have consistently failed to observe classical Na+ action potentials in 
{\em C. elegans} neurons~\cite{Goodman}.  
The failure to produce the stereotypical spike train dynamics normally associated with neuronal activity actually allow our model with graded electrical interaction to be more analogous with
observed activity in physical systems such as power grids~\cite{Mezic11}, thus broadening the scope of the work and its potential for impact in the physical sciences.

 \begin{figure}[t]
 \includegraphics[width=1.0\columnwidth]{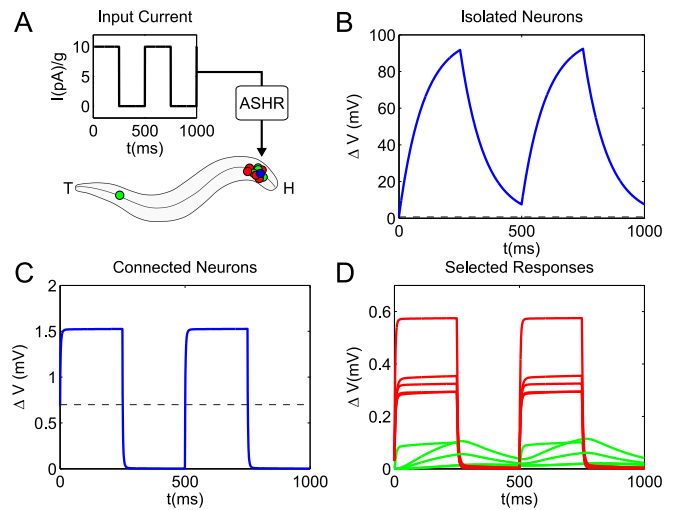}%
 \caption{\label{fig:model}
\textbf{A.} To demonstrate the behavior of the model, a square wave current input was injected into the neurons ASHR (units are pA divided by conductance constant $g=100\textrm{pS}$). Location of PLM neurons along the body of the worm are indicated by blue dots, and the location of 20 selected responding neurons are indicated by red dots.
\textbf{B.} Resulting membrane voltage displacement from equilibrium in ASHR when isolated (i.e. no connections).
\textbf{C.} Resulting membrane voltage displacement in ASHR when connected to the neuron network.
Note the two order of magnitude reduction in peak voltage.
\textbf{D.} Voltage responses of the 5 neurons most active in the presence of input current (in red) and the 5 neurons most active in the absence of input current (in green). 
The ASHR neuron is chosen so as to illustrate a ubiquitous phenomena: downstream 
neurons do not necessarily get entrained to the time-response of the stimulated neuron.
}
 \end{figure}

\subsection{Single-Compartment Membrane Model}

A model must be constructed for the graded response of neurons. Fortunately, it has been observed that many neurons in \emph{C. elegans} are effectively isopotential, such that we can use the membrane voltage as a state variable for network 
simulations~\cite{Goodman}. The time evolution of neuron $i$'s membrane potential, $V_i$, is therefore given by the single-compartment membrane equation~\cite{Wicks}:
\begin{equation}\label{eq:dvdt}
C\dot{V_i} = -G^{c} (V_i-E_{cell})-I_i^{Gap}(\vec{\textbf{V}})-I_i^{Syn}(\vec{\textbf{V}})+I_i^{Ext}
\end{equation}
$C$ is the whole-cell membrane capacitance, $G^{c}$ is the membrane leakage conductance  and  $E_{cell}$ is the leakage potential. The external input current is given by $I_i^{Ext}$, while neural interaction via gap junctions and synapses is modeled by input currents  $I_i^{Gap}(\vec{\textbf{V}})
$ (gap) and $I_i^{Syn}(\vec{\textbf{V}})$ (synaptic).  Their equations
are:
\begin{align}\label{eq:gaps1}
&I_i^{Gap} = \sum\limits_j G_{ij}^g(V_i-V_j)\\
&I_i^{Syn} = \sum\limits_j G_{ij}^{s} s_j(V_i-E_j)
\end{align}
Gap junctions are taken as ohmic resistances connecting each neuron where $G_{ij}^g$ is the total conductivity of the gap junctions between $i$ and $j$. Synaptic current is proportional to the displacement from reversal potentials $E_j$. $G_{ij}^s$ is the maximum total conductivity of synapses to $i$ from $j$, modulated by the synaptic activity variable $s_i$, which is governed by:
\begin{equation}\label{eq:synapticvar}
\dot{s_i} = a_r\phi(v_i;\kappa,V_{th})(1-s_i)-a_ds_i
\end{equation}
where $a_r$ and $a_d$ correspond to the synaptic activity's rise and decay time, and $\phi$ is the sigmoid function $\phi(v_i;\kappa,V_{th})=1/(1+\exp({-\beta(V_i-V_{th}})))$.

\subsection{Parameters}

While the precise parameter values of each connection are unknown, we assume reasonable values as previously considered in the literature~\cite{Varshney,Wicks}.    We assume each individual gap junction and synapse has approximately the same conductance, roughly $g=$100pS~\cite{Varshney}. Each cell has a smaller membrane conductance (taken as 10pS) and a membrane capacitance of about $C_i=1\textrm{pF}$~\cite{Varshney}. Leakage potentials are all taken as $E_c=-35\textrm{mV}$~\cite{Wicks}. Reversal potentials $E_j$ are 0mV for excitatory synapses and -45mV for inhibitory synapses~\cite{Wicks}. For the synaptic variable, we choose $a_r=1$, $a_d=5$, and define the width of the sigmoid by $\beta=0.125\textrm{mV}^{-1}$~\cite{Wicks}. $V_{th}$ is found by imposing that the synaptic activation $\phi=1/2$ at equilibrium~\cite{Wicks}.

The directionality of the connections (i.e., inhibitory or excitatory) is estimated by the rough approximation that putative GABAergic neurons are inhibitory, while cholinergic and glutamatergic neurons are excitatory (as in~\cite{Varshney}).
This estimation of parameter values captures robust responses in the network dynamics and excludes from the simulation any responses which depend on more precise details of the network.
The network that we simulate consists of 279 somatic neurons, where we exclude the 20 pharyngeal neurons and 3 additional neurons which make no synaptic connections, as in~\cite{Varshney}. To validate the simulation and the choice of parameters we tested for robustness by perturbing ($\pm$20\%) individual connection strengths and each neuron's parameters, showing that dynamic functionality persists.

\section{Analysis of Simulated Dynamics}

There are many ways to test the validity of the {\em C. elegans} model.  Given the numerous stimuli response experiments~\cite{Sharad,Aravi09,Aravi12}, we can simply select a neuron of interest and interrogate the downstream neuronal response.  For instance, the PLM neurons (PLML/R) are posterior touch mechanoreceptors. Activation of PLM by tail-touch causes a worm to move forward or, if already moving forward, to accelerate~\cite{Chalfie}. Thus stimulating these neurons should produce a downstream time-dependent neural-response resulting in a motorneuron response consistent with forward motion. Figure~\ref{fig:PLMdiagram} illustrates a schematic for this neuro-sensory cascade from sensory activation by stimulation of the sensory neuron PLM that excites the motor-neurons associated with forward motion~\cite{Aravi09}. Characterizing such neural pathways are the key objective in this study.
 
 \begin{figure}[t]
 \includegraphics[width=1.0\columnwidth]{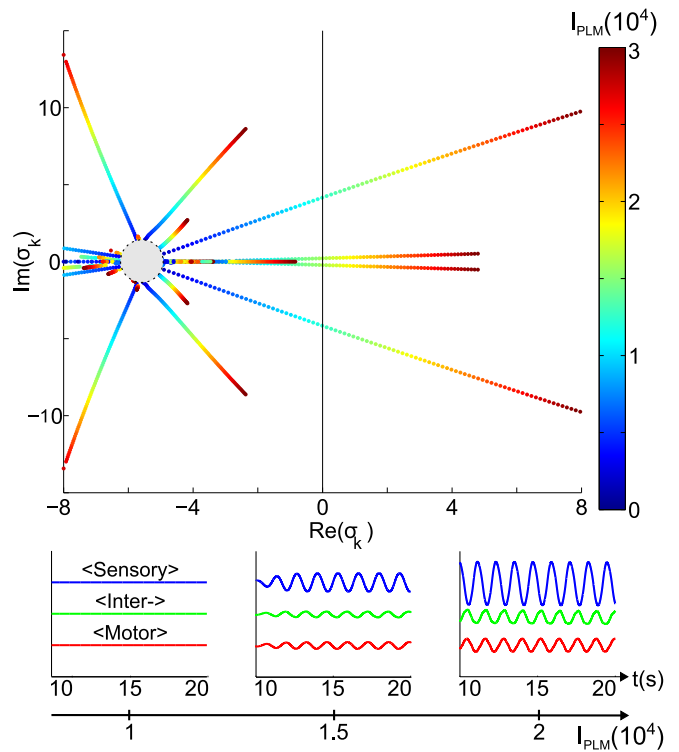}%
 \caption{\label{fig:Jeigs}
Jacobian eigenvalue spectrum as a function of PLM input amplitude. Input units are current normalized by conductance constant $g$. At inputs around $1\times10^4$, the system goes through a Hopf bifurcation and oscillatory motion results. Traces of average sensory, inter- and motor neuron voltage displacements from equilibrium are shown to illustrate this.}
 \end{figure}

A characteristic example of simulated full-network dynamics can be seen in Figure~\ref{fig:model}, in which the polymodal nociceptive neuron ASHR\cite{WormAtlas} is stimulated with a constant periodic input current. The system starts out in equilibrium before the excitation, and the voltages plotted are neuronal displacements from their equilibrium values. Panel B shows the response an unconnected neuron, whereas panel C shows the response of ASHR when connected to the network. Note that the characteristic timescale of the response changes due to the presence of connections. In panel D, the voltage responses are plotted for the 5 neurons which respond most strongly when input current is present, and for the 5 which respond most strongly when it is not. This illustrates that downstream neuron responses are not necessarily entrained to the stimulus, but may respond through different temporal modes.

 \begin{figure*}[t]
 \includegraphics[width=0.7\textwidth]{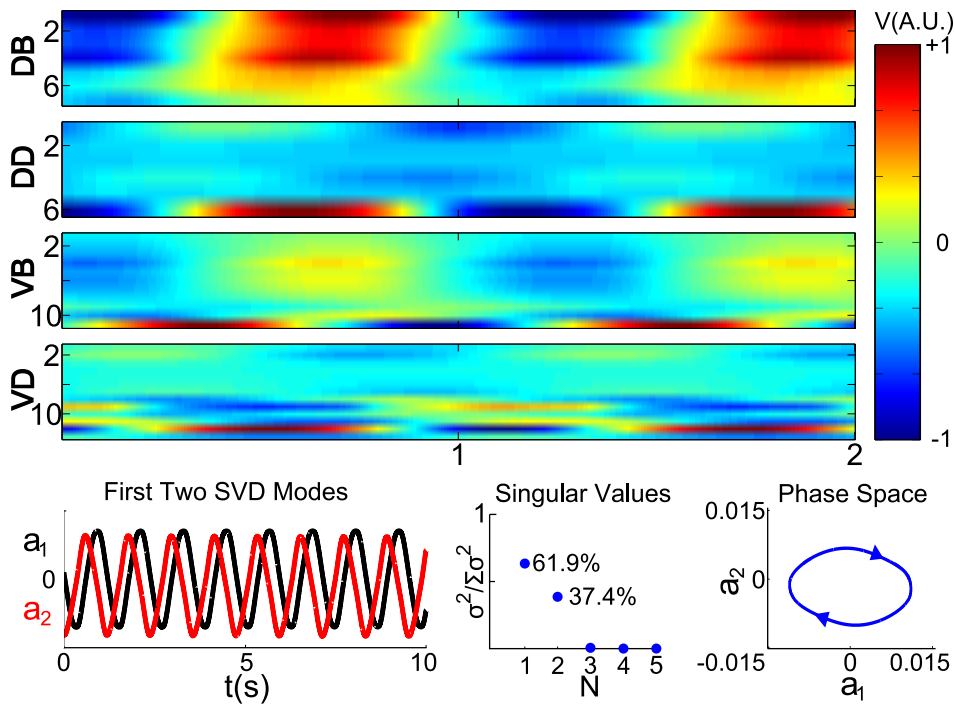}%
 \caption{\label{fig:DAVD}
Resultant dynamics for a constant input $2\times10^4$ into PLM (with this input, the system is within an oscillatory regime). Raster plots of the voltage responses within the forward motorneurons are shown at top. The time evolution of the SVD modes along with their singular values are shown on below, establishing that the system response is indeed dominated by two modes. The trajectory of these modes in phase space correspond to a two-mode swimmer.
}
 \end{figure*}

\subsection{Low-Dimensional Bifurcations}

Behaviorally, crawling is known to be dominated by a two-mode stroke motion~\cite{Bialek}, i.e. the so-called eigenworm motion.  Thus the motor-neuron response to PLM stimulation should produce a two-mode dominance in accordance with the eigenworm behavior given that the motor responses control muscle contraction~\cite{Aravi09}. We therefore intuitively anticipate that a constant input of sufficient strength, corresponding to sensory stimulus, should be able to drive two-mode oscillatory behavior in the forward motion motorneurons. To test if this is qualitatively captured by our model, we first seek oscillatory solutions by calculating the Jacobian matrix at equilibrium and looking for eigenvalues with positive real parts.

With zero external input, all Jacobian eigenvalues have a negative real part and the system is stable. However, eigenvalues with a positive real part are seen to exist for sufficiently high constant input amplitudes. Figure~\ref{fig:Jeigs} shows the Jacobian spectrum as a function of PLM input amplitude. At certain threshold values, the system goes through Hopf bifurcations and oscillatory modes arise. The average voltage displacement within each neuron class is shown on the right of the figure, illustrating this.

\subsection{Singular Value Decomposition of PLM response}

To obtain the modes that the motor neurons exhibit we collected time snapshots of motor-neuron voltages $\textbf{V}_M(t)$ into a matrix and computed the  singular value decomposition:
\begin{equation}\label{eq:snapshots}
V =
\begin{bmatrix} \textbf{V}_M(t_0) & \textbf{V}_M(t_1) & \hdots & \\
\end{bmatrix}=\textrm{P}\cdot \Sigma \cdot \textrm{Q}^T 
\end{equation}
The columns of matrix P (the vectors $\mathbf{P_i}$) are the principal orthogonal components, which are weighted by the diagonal elements in $\Sigma$ (the singular values $\sigma_i$). Decomposition of the voltage onto
these provides the dynamical coefficients $a_k(t)$:  
\begin{equation*}\label{eq:PODmodes}
\textbf{V}(t) = \sum_{k=1}^{N} a_k(t)\textbf{P}_k  
\end{equation*}

Figure~\ref{fig:DAVD} shows the time-dynamics of the motor neurons given constant PLM stimulation consistent with tail-touch at input amplitudes above the Hopf bifurcation level.  As shown, there are two dominant response modes that produce periodic, laterally out-of-phase, voltage activity.

The analysis as shown in the bottom row of Figure~\ref{fig:DAVD} confirms that the motor activity is dominated by two time-dependent response modes (with the first and second modes possessing 61.86\% and 37.36\% of the energy respectively). Their dynamics are periodic and similar to physiological~\cite{Bialek}  and behavioral~\cite{Sharad} studies that find
low-number of modes that determine the motion (specifically, there are two dominant oscillatory modes which move through their phase space in a ring around the origin).
Thus the model  produces a proxy for this behavior through analyzing motor responses, although it
does not produce {\em directly} the behavioral response. 

\section{Ablation}
Experimental ablation studies~\cite{Chalfie} have observed that the ablation of the densely-connected AVB interneurons destroys the worm's ability to perform forward motion, whereas ablation of the similarly densely-connected AVA interneurons preserves it (affecting instead the ability of the worm to perform backwards motion). If our model's PLM response modes do indeed serve as a proxy for this behavioral response, they should be similarly affected by such network modifications.

\begin{figure}
 \includegraphics[width=0.9\columnwidth]{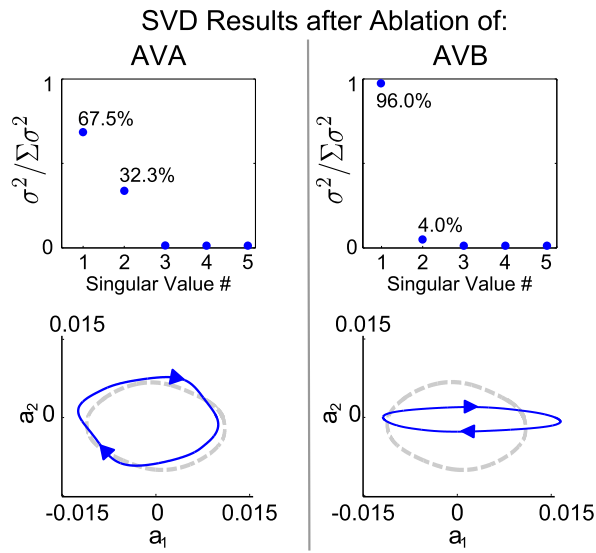}%
 \caption{\label{fig:Ablate}
Singular value distributions when the analysis of Figure~\ref{fig:DAVD} is repeated with the removal of the AVA or AVB interneurons from the network. Experimental studies~\cite{Chalfie} show that the ability to perform forward motion is destroyed with the removal of AVB, but preserved under the ablation of AVA. The second row shows the new trajectories in phase space after these ablations (where the dashed gray line is the healthy trajectory, for comparison). Note that AVA ablation does distort the trajectory, but does so less than does the ablation of AVB. Within our model, the two-mode response in the forward-motion motorneurons is affected in the same way by these ablations (AVB ablation destroys two-mode dominance, AVA ablation preserves it). This further suggests that the simulated neural modes serve as proxies for forward motion.
}
 \end{figure}

We explore the effect of ablation upon our response modes by removing the AVA/AVB interneurons from the network and repeating the analysis of Figure~\ref{fig:DAVD}. Specifically, the neurons AVAL/AVAR (or AVBL/AVBR) were removed from the network, the dynamics in response to an identical constant PLM input were simulated, and the SVD was calculated. The resultant singular value distributions for these ablations is shown in Figure~\ref{fig:Ablate}, which shows that the two-mode dominance is destroyed with the removal of AVB but remains intact with the removal of AVA. This serves as another confirmation that the response modes correspond to the experimental foward-motion modes.

\begin{figure*}
 \includegraphics[width=0.9\textwidth]{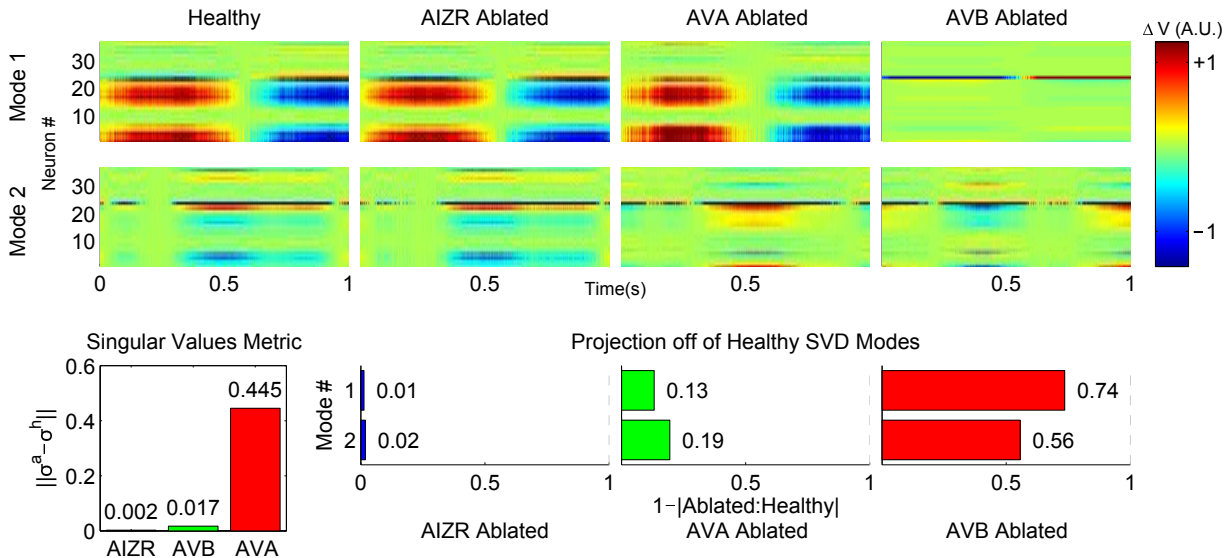}%
 \caption{\label{fig:AblateMetrics}
Modes after ablation of neurons compared to healthy modes. The raster plots at top show the first and second SVD modes of forward-motion motorneuron activity for the healthy case (full network, no ablations), along with the cases where AIZR, AVA and AVB are ablated. Phase-matched one-second intervals of each mode are shown. On the bottom row the metrics of Equations~\ref{eq:metSV} and~\ref{eq:metFrob} are shown for the ablation modes compared to the healthy modes. 
}
\end{figure*}

\subsection{Change in Response Modes}

The top rows of Figure~\ref{fig:AblateMetrics} show the dynamics of the first two SVD modes (i.e. the time evolution of $a_k(t)\cdot\mathbf{P_k}$ for modes $k=1,2$) before any ablations (for the "healthy" system) and after ablation of AVA, AVB, and AIZR (the latter being chosen because, experimentally, the ablation of AIZR does not inhibit forward motion\cite{WormAtlas}). The 37 neurons selected are the forward-motion motorneurons (those belonging to classes DB, DD, VB and VD, as in Figure~\ref{fig:DAVD}). A time interval of 1 second was selected from each simulation such that the first modes of all cases were maximally in-phase. Note that the same observation as before can be made when qualitatively comparing the structures of the modes of the healthy, AVA ablated and AVB ablated case: when the AVA interneurons are ablated, the structure of the modes appears slightly altered, but similar, whereas ablation of AVB destroys the dominant mode.

\subsection{Quantification of Response Similarity}

To quantify the effect of ablations on the response modes and their dynamics we introduce two metrics. The first metric measures the similarity of the singular values by computing the $l^2$-norm between the ablated and healthy distributions. For ablated singular values $\sigma^a$ and healthy singular values $\sigma^h$, we compute:

\begin{equation}\label{eq:metSV}
||\sigma^a-\sigma^h||=\sqrt{\sum_{i=0}^N(\sigma^a_i-\sigma^h_i)}
\end{equation}

The second metric computes the similarity between the mode dynamics. We take the one-second dynamics segments from Figure~\ref{fig:AblateMetrics} (labeled here as matrix $\mathrm{H}$ for the healthy modes and $\mathrm{A}$ for the ablated), and we compute the absolute value of their Frobenius product:

\begin{equation}\label{eq:metFrob}
|\mathrm{A} : \mathrm{H}| = \left|\sum_{i,j}\mathrm{A}_{ij}\cdot\mathrm{H}_{ij}\right|
\end{equation}

where all matrices have been normalized to have a Frobenius norm of one.

Figure~\ref{fig:AblateMetrics}, on the bottom row, shows the values of these metrics for the ablations of AVA, AVB and AIZR. By these metrics, ablation of AIZR does not affect the network's response to this stimulus, ablation of AVA affects only slightly the network's response, and ablation of AVB destroys the functionality of the network in response to a PLM stimulus. This suggests that such analysis can be used to computationally classify the roles played by specific neurons in the response of the network to given stimuli. Hence our model provides a computational framework in which to computationally classify, from no prior knowledge, the neural subnetworks responsible for behavioral responses to stimuli.

\section{Conclusion}

In conclusion, we have developed a neuro-sensory integration model of the {\em C. elegans} nematode which describes the nonlinear, time-dependent, network voltage conductances.  In our computational model, the entire 302 neuron network of sensory-, inter- and motor-neurons are dynamically coupled with the best available biophysical connectome data to date. In the specific application of the tail mechanosensory neuron PLM stimulation, a complete neuro-sensory integration of this specific stimulus pathway is discovered whereby sensory information translates to downstream motor responses that are responsible for behavioral actions, in this case a two-mode swimmer dynamics.
In theoretical terms, the input stimulus robustly induces a Hopf bifurcation in the network.
Thus a low-dimensional bifurcation, which is ultimately responsible for behavior, is inscribed in the underlying network structure.

With the abundant current and on-going biophysical experiments on individual neuron stimulation in {\em C. elegans} (through opto-genetics, for instance), the current model presents a significant step forward in providing a theoretical platform to more accurately understand neuro-sensory encoding, processing, and integration. 
Specifically, we construct a biophysically inspired computational model and demonstrate that the
underlying low-dimensional bifurcations of the network drive
neural voltage modes which are responsible for low-dimensional movement and behavior.
These neural modes can be linked to behavioral responses via comparison with the experimentally observed behavioral effects of neural network modification. Thus our model allows for the identification and characterization of behavioral responses which are encoded within low-dimensional bifurcation structures in the network. The identification of such bifurcation-encoded responses within the network allows for computational classification of neurons into the subnetworks responsible for those responses. 

Our study thus allows one to study the structure and robustness of networks of voltage conductances for producing prescribed responses. 
More broadly, understanding the {\em C. elegans} model organism may help produce and promote bio-inspired network designs in other fields of scientific applications given the observed robust nature of such architectures.
This study promotes a viewpoint of the broader potential for understanding what can be
gained in modeling physical systems whose dynamics are driven by network connectivity and
nonlinear dynamical systems.  This can lead to bio-inspired design, quantification and engineering principles capable of producing robust functionality.

\begin{acknowledgments}
We are especially indebted to Sharad Ramanathan 
and Aravi Samuel for insight concerning the role of sensory 
and inter-neurons and {\em C. elegans} behavior.
J. N. Kutz acknowledges support
from the National Science Foundation (DMS-1007621).
\end{acknowledgments}

\bibliography{ce_pre}

\end{document}